\documentclass[journal]{IEEEtran}

\usepackage{amsmath,epsfig,amssymb,verbatim}
\usepackage{cite}
\usepackage{graphicx}
\usepackage{bm,bbm}
\usepackage{lipsum}
\usepackage{subfigure}

\newtheorem{Theorem}{Theorem}

\newcommand{\e}{\epsilon}

\newcommand{\lbd}{\lambda}

\newcommand{\E}{\mathbb{E}}

\begin{document}

\title{Whittle Index Based Scheduling Policy for Minimizing the Cost of Age of Information}

\author{Zhifeng Tang,~\IEEEmembership{Student Member,~IEEE,}
Zhuo Sun,~\IEEEmembership{Member,~IEEE,}
Nan Yang,~\IEEEmembership{Senior Member,~IEEE,}\\
and Xiangyun Zhou,~\IEEEmembership{Senior Member,~IEEE}\vspace{-1em}

\thanks{This work was funded by the Australian Research Council Discovery Project (DP180104062).}
\thanks{Z. Tang, N. Yang, and X. Zhou are with the School of Engineering, Australian National University, Canberra, ACT 2600, Australia (Email: \{zhifeng.tang, nan.yang, xiangyun.zhou\}@anu.edu.au).

Z. Sun is with the School of Computer Science, Northwestern Polytechnical University, Xi’an, Shaanxi 710072, China. She was with the School of Engineering, Australian National University, Canberra, ACT 2600, Australia (Email: zsun@nwpu.edu.cn).}}


\maketitle

\begin{abstract}
We design a new scheduling policy to minimize the general non-decreasing cost function of age of information (AoI) in a multiuser system. In this system, the base station stochastically generates time-sensitive packets and transmits them to corresponding user equipments via an unreliable channel. We first formulate the transmission scheduling problem as an average cost constrained Markov decision process problem. Through introducing the service charge, we derive the closed-form expression for the Whittle index, based on which we design the scheduling policy. Using numerical results, we demonstrate the performance gain of our designed scheduling policy compared to the existing policies, such as the optimal policy, the on-demand Whittle index policy, and the age greedy policy.

\end{abstract}


\IEEEpeerreviewmaketitle

\section{Introduction}\label{sec:introduction}

Ultra-reliable and low-latency communication (URLLC) has been acknowledged as one of the enabling communication paradigms for the fifth generation (5G) networks \cite{Li2019}. As many real-time URLLC applications emerge \cite{Simsek2016}, e.g., intelligent transportation and factory automation, the timeliness of information becomes increasingly critical. In order to fully characterize the freshness of delivered information, a new performance metric -- Age of Information (AoI) -- was proposed \cite{Kaul2011}. In particular, the AoI is defined as the time elapsed since the latest successfully received information was generated by the transmitter, which captures both the latency and the generation time of each information update.

Since being introduced in \cite{Kaul2011}, the concept of AoI has reaped a wide range of attention. Starting from analyzing the AoI performance in \cite{Yates2012,Kaul2012,WangGC2019,Chen2020,Wang2020IoT}, some transmission policies were designed to effectively improve the AoI performance \cite{Jiang2019,Wang2019a,Kadota2018,Tang2020jsac,abd2020,Kadota2016,Hsu2018,Sun2019A}. Among them, the transmission scheduling policy was optimized to minimize the average AoI for multiple-source systems in \cite{Kadota2016,Hsu2018,Sun2019A}. In particular, \cite{Kadota2016} proposed a Whittle index based scheduling policy to minimize the average AoI, where both a deterministic packet generation model and a reliable link were considered. \cite{Hsu2018} extended it to a stochastic packet generation model, while considering a system without buffer. In \cite{Sun2019A}, buffers were introduced and a Whittle index based scheduling policy was designed to minimize the average AoI under an unreliable link.

In the aforementioned studies, the performance metric employed to design the scheduling policy is the average AoI. Recently, the general function of the AoI has been introduced as a natural extension to the average AoI \cite{sun2019,Kosta2020,Klugel2019,tripathi2019}, which characterizes how the level of dissatisfaction depends on data staleness. In \cite{sun2019}, the weighted sum of the AoI of all sources was adopted as the performance metric to optimize the scheduling policy. Then the non-linear functions of the AoI were proposed in \cite{Kosta2020,Klugel2019,tripathi2019}. In \cite{Kosta2020}, the average cost of AoI was derived for three sample functions in an M/M/1 queue model with a first-come-first-served (FCFS) queue discipline. In \cite{Klugel2019}, a threshold based scheduling policy was proposed to optimize a general average cost of AoI in network control systems. Considering a general non-decreasing functions of the AoI, \cite{tripathi2019} designed the Whittle index based scheduling policy. However, only the deterministic packet generation model was considered in \cite{tripathi2019}, while the stochastic packet generation model remains unexplored. This motivates our work.

In this paper, we design a new scheduling policy that minimizes the \textit{general non-deceasing cost function of AoI} in a multiuser system with stochastic packet arrivals and unreliable channels. Comparing with \cite{Tang2020jsac} and \cite{abd2020}, we analyze the impact of stochastic packet arrivals on the scheduling policy design to minimize the \textit{general non-deceasing cost function of AoI}, instead of AoI itself. In particular, we first derive the closed-form expression for Whittle index and establish the corresponding indexability. Comparing with \cite{tripathi2019}, which considered deterministic packet generation, the index derivation is more challenging since we analyze the impact of the stochastic packet arrival on the scheduling policy design. By considering stochastic packet generation, the queuing delay forms the second-dimension of the system state. The one-dimensional analysis in \cite{tripathi2019} cannot be applied to the two-dimensional system. 
Based on the closed-form expression for Whittle index, we propose a scheduling policy to minimize the \textit{general non-deceasing cost function of AoI}. Aided by numerical results, we show that our proposed policy achieves profound AoI performance improvement compared to the existing policies, for given non-decreasing functions. We also show that our proposed policy achieves a larger AoI performance gain in the heterogeneous case where user equipments (UEs) have different packet generation probabilities than the homogeneous case where UEs have the same packet generation probability.


\section{System Model and Problem Formulation}\label{sec:system}

\begin{figure}[!t]
    \centering
    \includegraphics[width=0.7\columnwidth]{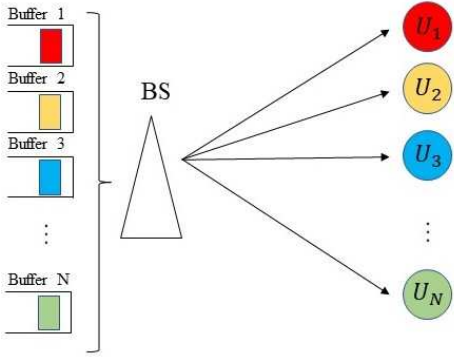}
    \vspace{-1em}
    \caption{Illustration of our considered time-slotted system where BS transmits the packets to $N$ UEs.}\label{fig:system_model}
    \vspace{-2em}
\end{figure}

We consider a time-slotted multiuser system, as depicted in Fig.~\ref{fig:system_model}, where the base station (BS) transmits packets to $N$ different UEs, denoted by $\Phi_{U} =\{U_1, U_2, \cdots, U_N\}$. We assume that there are $N$ buffers at the BS, each of which corresponds to one UE. At the beginning of each time slot, the BS generates the packet of $U_n$ with a probability $\lambda_n$, $n \in \{1,2,\ldots,N\}$\footnote{We assume that the packet generation process is independent but not identical among UEs.}. When the packet of $U_n$ is generated, it is stored in its corresponding buffer. We further assume that each buffer stores one packet such that the newly generated packet replaces the previous one in the buffer.

We assume that the BS schedules at most one packet transmission in each time slot, and hence, only the packet of one UE can be transmitted at a time. We denote a binary variable $u_{n}(t)=\{0,1\}$ as the scheduling indicator of $U_n$ during time slot $t$. If $U_n$ is scheduled for transmission during time slot $t$, $u_{n}(t)=1$. Otherwise, $u_{n}(t)=0$. Thus, we obtain
\begin{align}\label{eq:constrain}
\setlength{\abovedisplayskip}{1pt}
\setlength{\belowdisplayskip}{1pt}
\sum_{n=1}^{N}u_n(t)\leq 1, \forall t.
\end{align}
We assume that during each time slot, packet transmission occurs prior to new packet generation. Thus, only the packets generated in previous time slots can be transmitted by the BS.

By considering practical signal propagation between the BS and UEs, the transmitted packets may not be successfully detected by UEs. We denote a binary variable $\delta_n(t) = \{0,1\}$ as an indicator of whether or not $U_n$ successfully detects its packet transmitted by the BS during time slot $t$. If the packet is successfully detected by $U_{n}$ during time slot $t$, $\delta_n(t)=1$. Otherwise, $\delta_n(t)=0$. The transmission error probability of $U_n$ is denoted by $\epsilon_{n}$, i.e., $\mathrm{Pr}(\delta_n(t)\!=\!0)=\epsilon_n$. When $U_{n}$ successfully detects a packet, it immediately sends back an acknowledgement (ACK) to the BS via an error-free control channel. The feedback overhead is assumed to be negligible as the ACK length is much smaller than the packet length. Once an ACK is received from $U_n$, the BS empties the corresponding buffer by dropping the successfully received packet. Otherwise, this packet is held in the buffer.

We denote $V_{n}(t)$ as the generation time of the last successfully detected packet at $U_n$ during time slot $t$. Then, the AoI of $U_n$ during time slot $t$ is given by
\begin{align}\label{eq:secondeq}
h_n(t) = t - V_n(t).
\end{align}
Let us define $W_{n}(t)$ as the generation time of the newest generated packet of $U_n$. Until time slot $t$, the queuing delay of this packet is computed as $a_{n}(t) = t - W_n(t)$. The duration between the generation time of the last successfully detected packet and that of the newest generated packet is $d_n(t) = W_n(t) - V_n(t)$. Then, we express the AoI of $U_n$ as $h_{n}(t) = a_{n}(t)+d_{n}(t)$. Thus, the evolution of the AoI of $U_n$ is given by
\begin{align}\label{eq:AoIevolution}
h_n(t+1)=\left\{
\begin{aligned}
&a_n(t)+1, &&\textrm{if } u_n(t)\delta_n(t) = 1,\\
&h_n(t)+1, && \textrm{otherwise}.
\end{aligned}
\right.
\end{align}
It is noted that when the BS successfully transmits a packet to $U_n$ but does not generate a new packet, this successfully transmitted packet is considered as the newest generated packet. When this happens, we have $d_n(t)=0$ and $h_n(t)=a_n(t)$.

Based on \eqref{eq:secondeq}, we define the average cost of AoI as
\begin{equation}\label{eq:systemAoI}
\Xi=\lim_{T\rightarrow\infty}\frac{1}{TN}
\sum_{t=1}^{T}\sum_{n=1}^{N}v(h_{n}(t)),
\end{equation}
where $v(h_n(t))$ is a general non-decreasing function of AoI, characterizing the importance level of packet freshness. We observe from \eqref{eq:AoIevolution} that the AoI is determined by the adopted transmission scheduling policy, i.e., $u_n(t)$. Thus, we can optimize the transmission scheduling policy to minimize the average cost of AoI in \eqref{eq:systemAoI}. The optimization problem is formulated as
\begin{align}\label{eq:MDPpro}
\Xi_{c}=\min_{c\in\mathcal{C}}&\lim_{T\rightarrow \infty}\frac{1}{NT}\sum_{t=1}^T\sum_{n=1}^N v( h_{n}(t)),\notag\\
&\textrm{s.t.}\quad \sum_{n=1}^N u_n(t) \leq 1, \forall t,
\end{align}
where $\mathcal{C}$ denotes the set of all potential transmission scheduling policies. In fact, this optimization problem is an infinite time horizon average cost constrained Markov decision process (CMDP) problem, with the state space $\{(a_1(t),d_1(t)),\cdots,(a_N(t),d_N(t))\}$. We note that, due to the countably infinite state space of \eqref{eq:MDPpro}, it is infeasible to obtain the optimal scheduling policy by using conventional CMDP methods, e.g., value iteration and policy iteration. To address this infeasibility, we exploit the Whittle index method to obtain a sub-optimal scheduling policy. In particular, we introduce a constant service charge $m$ as the minimum service charge of the system to schedule transmission \cite{whittle_1988}. Based on this introduction, we decouple the original CMDP problem into sub-problems with a smaller state space, where each sub-problem only involves one UE. Therefore, in order to solve the problem in \eqref{eq:MDPpro}, we optimize the average cost of AoI of each UE individually by a Lagrange function with a multiplier $m$, given by
\begin{align}\label{eq:MDPpro_decouple}
J^{\ast}=\min_{u(t)\in\{0,1\}}\frac{1}{T}\sum_{t=1}^T \E\Big[&v\big(a(t)+d(t)\notag\\
&\times(1-u(t)\delta(t))\big)+mu(t)\Big].
\end{align}
Since the AoI analysis for each UE is identical, we omit the index $n$ from $a_n(t)$, $d_n(t)$, $u_n(t)$, and $\delta_n(t)$ in \eqref{eq:MDPpro_decouple} to allow easy readability. Finally, to guarantee the existence of a finite average cost of AoI, we clarify that the cost function of AoI, $v(h)$, needs to satisfy $\sum_{k=0}^{\infty}\e^k v(k)<\infty$ and $\sum_{k=0}^{\infty}(1-\lbd)^k v(k)<\infty$. Otherwise, the average cost of AoI goes to infinity.

\section{Index Policies}\label{sec:index}

In this section, we derive the Whittle index and propose a Whittle index based scheduling policy. To derive the Whittle index, we first obtain the minimum average cost of AoI by solving \eqref{eq:MDPpro_decouple}. Here, we omit the time index $t$ and present the Bellman equation as
\begin{align}\label{eq:Bellmaneq}
J^{\ast}+f(a,d)=\min\left\{\mu_0(a,d),\mu_1(a,d)\right\},
\end{align}
where
\begin{align}
\mu_0(a,d)=&v(a\!+\!d)+\!\lambda f(1,a\!+\!d)\!+\!(1\!-\!\lambda) f(a\!+\!1,d),\\
\mu_1(a,d)=&m\!+\!\epsilon(v(a\!+\!d)\!+\!\lambda f(1,a\!+\!d)\!+\!(1\!-\!\lambda) f(a\!+\!1,d))\!\notag\\
+&(1\!-\!\epsilon)\left(v(a)\!+\!\lambda f(1,a)\!+\!(1\!-\!\lambda) \!f(a\!+\!1,0)\!\right),
\end{align}
and $f(a,d)$ is the differential cost-to-go function with $f(1,0)\! =\! 0$. We assume that given $a$, $f(a,d)$ in \eqref{eq:Bellmaneq} is non-decreasing with $d$, i.e., $f(a,0)\leq f(a,1)\leq f(a,2)\leq\cdots$. Based on this, the optimal policy in \eqref{eq:MDPpro_decouple} is proven to be threshold-based \cite{Bertsekas2000}, denoted by $c_{D}$. In particular, the action of state $(a,d)$ is to idle when $d<D_a$, and to schedule when $d> D_a$, where $D_a$ is the threshold and satisfies $D_1\leq D_2\leq\cdots\leq D_a\leq\cdots$. In addition, both idle and schedule actions are equally appealing for state $(a,D_a)$. Thus, the design of this threshold-based policy $c_{D}$ is equivalent to obtain the threshold $D_a$. We next derive the threshold $D_a$ in the following Theorem.

\begin{Theorem}\label{Theorem1}
For the threshold-based policy $c_D$, the threshold $D_a$ satisfies
\begin{align}\label{eq:thresholdD1}
\lbd\e&\omega(a+D_a)+\psi(a+D_a)-\e\theta(D_1+1)\notag\\
&=\frac{1}{D_1}\left(a\!+\!\frac{1}{\lbd}\!-\!1\right)\left(\frac{m}{1\!-\!\e}\!+\!\sum_{h=1}^{D_1}v(h)\right)\!-\!\sum_{h=1}^{a-1}v(h),
\end{align}
for $a< D_1$, and
\begin{align}\label{eq:thresholdlD2}
\psi(a\!+\!D_a)+\lbd\e\omega(a\!+\!D_a)=\frac{m}{1-\e}+\psi(a)+\lbd\e\omega(a),
\end{align}
for $a \geq D_1$, where $\theta(h)=\sum_{k=0}^{\infty} \e^k v(h+k)$, $\psi(\!h\!)\!=\!\sum_{k\!=\!0}^{\infty} (1\!-\!\lbd)^k v(h\!+\!k)$, and $\omega(\!h\!)\!=\!\sum_{k\!=\!1}^{\infty} \e^{k\!-\!1}\theta(\!h\!+\!k\!)$ if $\e\!=\!1\!-\!\lambda$; otherwise, $\omega(h)=\left(\psi(h)-\theta(h)\right)/(1-\lbd-\e)$.
\begin{IEEEproof}
See Appendix~\ref{Appendix:A}.
\end{IEEEproof}
\end{Theorem}

Based on Theorem~\ref{Theorem1}, we can derive the indexability of the threshold policy in Theorem~\ref{Theorem:Index}.
\begin{Theorem}\label{Theorem:Index}
Considering the decoupled model, the scheduling policy $c$ in Theorem \ref{Theorem1} is indexable.
\begin{IEEEproof}
When $m=0$, we obtain the thresholds $D_a$ in Theorem \ref{Theorem1} for all $a$ as zero and hence, the idle state space is empty. When $m$ goes to infinity, the thresholds $D_a$ go to infinity. Hence, the idle state space is the entire space. In addition, for any cost $m_1<m_2$, it is clear that any state $(a,d)$ is the idle state for $m_2$ if it is the idle state for $m_1$. Hence, the idle state space for $m_1$ is a subset of the idle state space for $m_2$.
\end{IEEEproof}

\end{Theorem}

The Whittle index is defined as the minimum auxiliary service charge to ensure that both the action of being scheduled and the action of being idle are equally appealing for the current state \cite{whittle_1988}. In other words, the Whittle index is obtained as the minimum $m$ to obtain $d = D_a$ for the state $(a,d)$. Based on this, we derive the Whittle index in Theorem \ref{Theorem:2}.
\begin{Theorem}\label{Theorem:2}
Let us consider one UE has a packet generation probability $\lambda$ and a transmission error probability $\epsilon$. When this UE is in the state $(a,d)$, its Whittle index is given by
\begin{align}\label{eq:Whittle}
&I_v(a,d,\lbd,\e)\notag\\
&=\begin{cases}
(1\!-\!\e)\Big(\!\lbd(1\!-\!\e)d\omega(d)\!-\!\sum\limits_{h\!=\!1}^{d}v(h) \Big),&\hspace{-5mm}\textrm{if}~a\!=\!1,\\
(\!1\!-\!\e\!)\Big(\!\lbd(\!1\!-\!\e\!)D_1\omega(D_1)\!-\!\sum\limits_{h\!=\!1}^{D_1}\!v(h) \!\Big),&\hspace{-5mm}\textrm{if}~2\!\leq\! a\!\leq\! D_1,\\
(\!1\!-\!\e)\!\big(\!\psi(\!a\!+\!d)\!-\!\psi(a)
\!+\!\lbd\e\!\left(\!\omega(\!a\!+\!d)\!-\!\omega(\!a)\!\right)\!\big),&\textrm{if}~a\!>\!D_1,
\end{cases}
\end{align}
where $D_1$ is the minimum positive number satisfying
\begin{align}\label{eq:WD1calcu}
\e\theta(D_1\!+\!1)+\lbd(1&-\e)\left(a\!+\!\frac{1}{\lbd}\!-\!1\right)\omega(D_1)\notag\\
=&\lbd\e\omega(a+d)+\psi(a+d)+\sum_{h=1}^{a-1}v(h).
\end{align}
\begin{IEEEproof}
See Appendix~\ref{Appendix:B}.
\end{IEEEproof}
\end{Theorem}

We remark that when considering the special case with a deterministic packet generation model, i.e., $\lambda=1$, the Whittle index in \eqref{eq:Whittle} coincides with the results in \cite{tripathi2019}. Furthermore, when considering the average cost of AoI being given by the average AoI, i.e., $v(h)=h$, the derived Whittle index in \eqref{eq:Whittle} matches the result in \cite{Sun2019A}. Therefore, the derived Whittle index in \eqref{eq:Whittle} is a general result for stochastic packet generation models and any cost function of AoI.

Based on the derived Whittle index in \eqref{eq:Whittle}, we propose the optimal transmission scheduling policy that minimizes the average cost of AoI. In particular, for each time slot, the BS schedules the UE with the highest values of Whittle index for transmission. This is because that for the UE with a larger Whittle index, the transmission of its packet makes more contributions to reducing the average cost of AoI in the system.

\section{Numerical Result and Discussion}\label{sec:numerical}

In this section, we present numerical results to demonstrate the effectiveness of our proposed Whittle index based scheduling policy in Section~\ref{sec:index}. For comparison, we first compare our scheduling policy with the optimal scheduling policy. We then employ the on-demand Whittle Index policy \cite{tripathi2019} and age greedy scheduling policy \cite{Sun2019A} as the benchmark policies. 

\begin{figure}[!t]
\centering
\includegraphics[width=0.8\columnwidth]{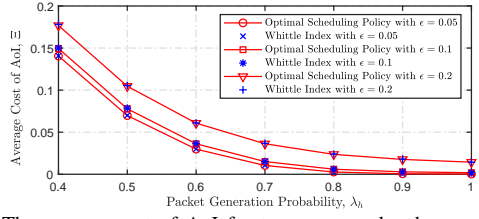}
\vspace{-1em}
\caption{The average cost of AoI for two users under the proposed Whittle index scheduling policy and an optimal scheduling policy.}\label{fig:OPT}
\vspace{-1em}
\end{figure}

\begin{figure}[!t]
\centering
\includegraphics[width=0.8\columnwidth]{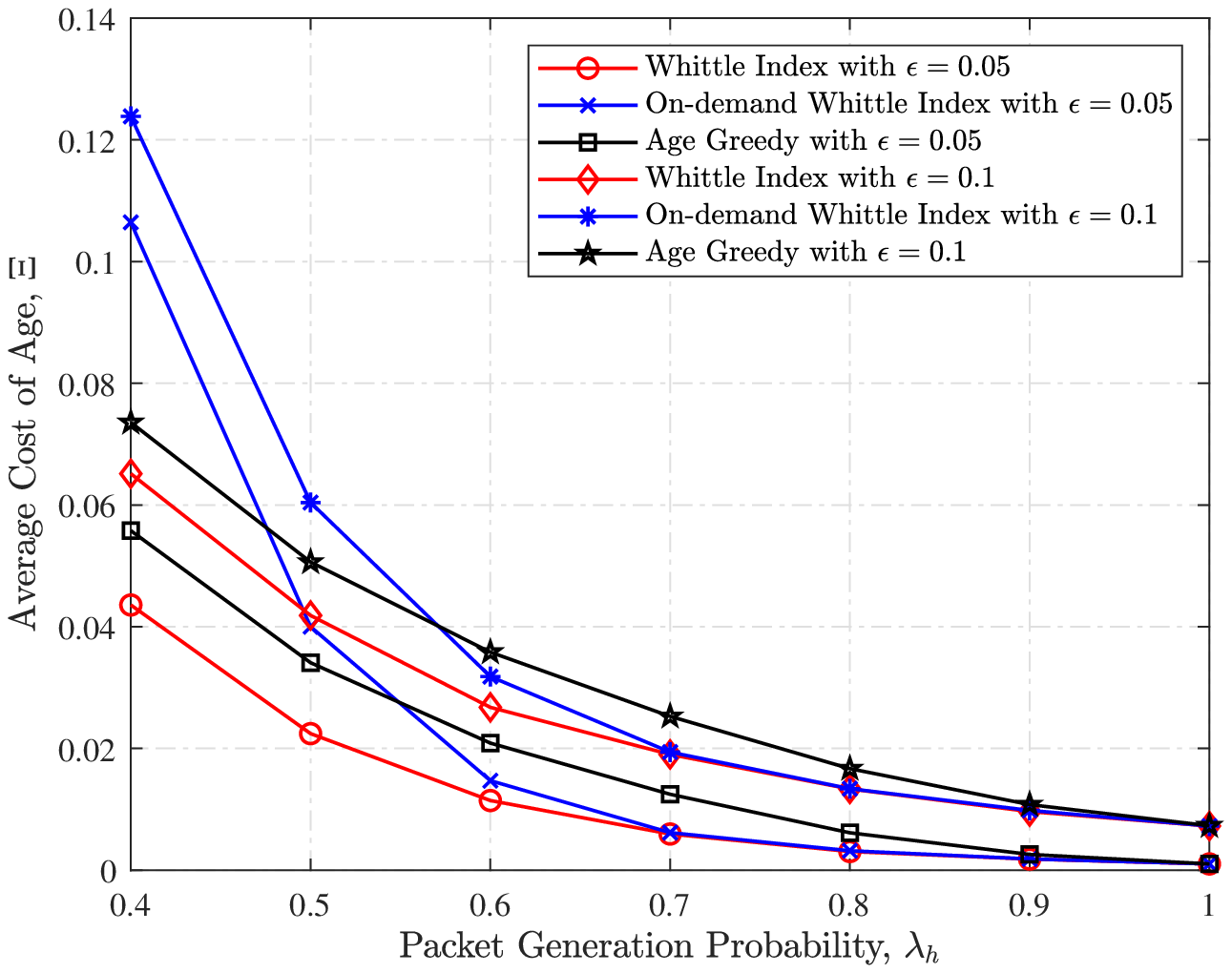}
\vspace{-1em}
\caption{The average cost of AoI versus the packet generation probability with $N = 6$.}\label{fig:2ps}
\vspace{-1.5em}
\end{figure}

\begin{figure}[!t]
\centering
\includegraphics[width=0.8\columnwidth]{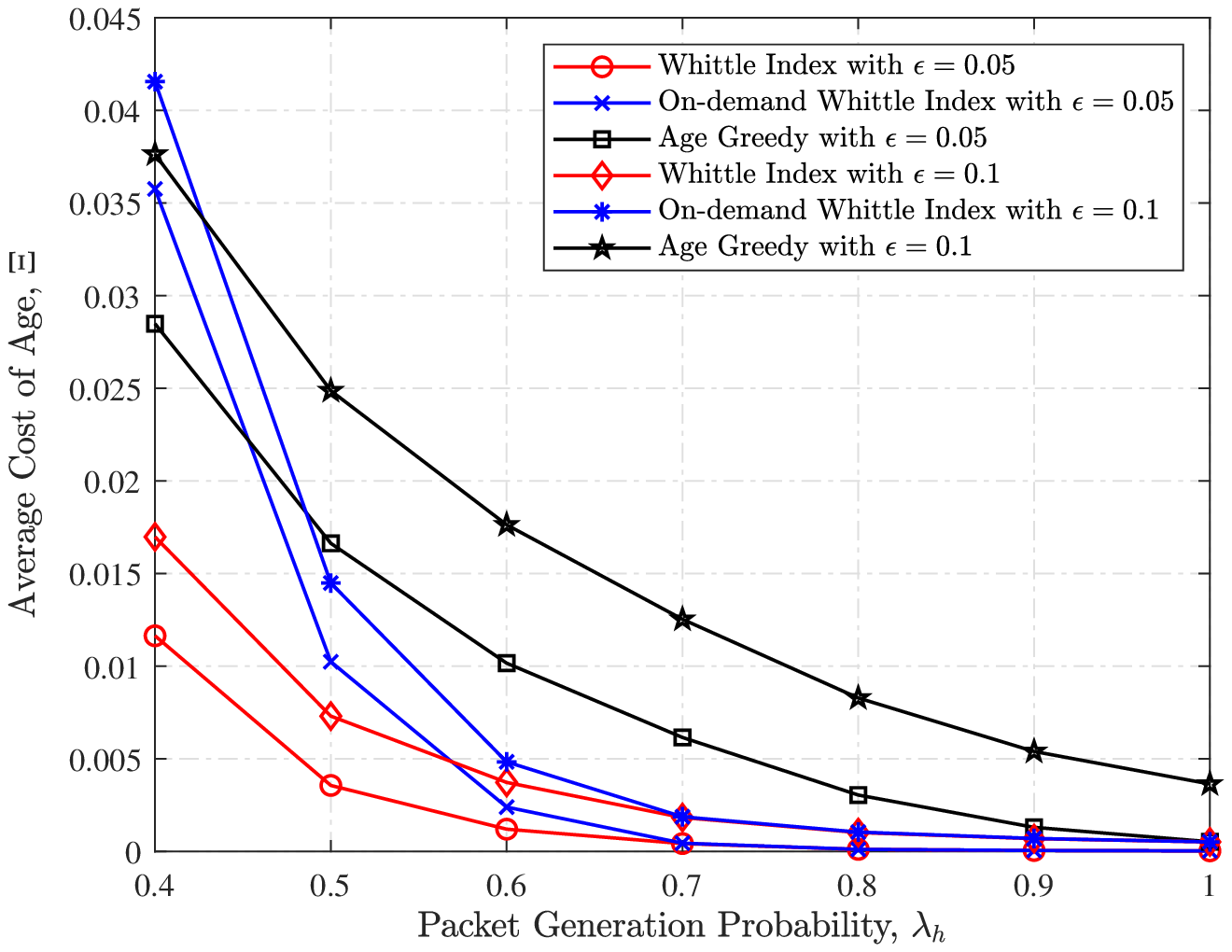}
\vspace{-1em}
\caption{The average cost of AoI versus the packet generation probability with $N = 6$.}\label{fig:3ps}
\vspace{-1.5em}
\end{figure}

In Fig.~\ref{fig:OPT}, we compare the proposed scheduling policy with the optimal scheduling policy for two users with an AoI violation function as the cost function such that $v(h)=1$ if $h\geq6$; otherwise, $v(h)=0$. It turns out that the designed scheduling policy almost achieves the minimum average cost of AoI, which shows the optimality of our designed scheduling policy.
    
Fig.~\ref{fig:2ps}. plots the average cost of AoI versus the packet generation probability, $\lambda$, for the homogeneous case where all UEs share the same packet generation pattern and the same transmission error probability, i.e, $\lambda_n\! = \!\lambda$, and $\epsilon_n \!=\! \epsilon$ for $n\!=\!1,2,\!\cdots\!,N$. We employ an AoI violation function as the cost function such that $v(h)=1$ if $h\geq10$; otherwise, $v(h)=0$. We observe that for all $\epsilon$, the designed scheduling policy achieves a lower average cost of AoI than the benchmark policies, especially for small $\lambda_h$, which shows the advantage of the designed scheduling policy over existing policies. This is due to the fact that the impact of packet freshness on the average cost of AoI increases when $\lambda$ decreases and our proposed policy design is forward-looking instead of myopic, i.e., it evaluates the long term payoff. 
    
Fig.~\ref{fig:3ps}. plots the average cost of AoI versus the packet generation probability, $\lambda$, for the heterogeneous case where half UEs have the same cost function such that $v(h)=1$ if $h\geq10$; otherwise, $v(h)=0$. The other half UEs have another cost function such that $v(h)=1$ if $h\geq15$; otherwise, $v(h)=0$. Given different cost functions among UEs, we observe that compared to the benchmark policies, our proposed policy achieves a larger AoI performance gain for the heterogeneous case than for the homogeneous case. This is due to the fact that UEs have different dissatisfaction levels of data staleness, which results in different impacts of UEs' successful packet transmission on the reduction in the average cost of AoI. Importantly, such impacts are addressed in our proposed policy.

\section{Conclusion}\label{sec:con}

This paper considered a multiuser system where a BS generates time-sensitive packets and transmits them to UEs under an unreliable channel. We employed the average cost of AoI as the metric to characterize the freshness of transmitted packets. By introducing a constant service charge, we derived the closed-form expression for the Whittle index. Based on this expression, we proposed a Whittle index based scheduling policy. Under this proposed scheduling policy, Whittle index reveals the value of each packet and the BS intends to send the most valuable packet to reduce the average cost of AoI. Using simulations, we showed that the performance gain of our proposed policy.

\begin{appendices}

\section{Proof for Theorem \ref{Theorem1}}\label{Appendix:A}

We first give the expression of $f(a,d)$ as

\begin{align}\label{eq:fad}
f(a,d)\!=\!\left\{
\begin{aligned}
&(\!a\!+\!d\!-\!1)J^{\ast}\!-\!\sum_{h=1}^{a\!+\!d\!-\!1}v(h)\;,\   \mbox{\textrm{if }$1\!\leq\! a\!+\!d\leq D_1$},\\
&\frac{m}{1\!-\!\epsilon}\!-\!\frac{J^{\ast}}{\lambda}\!+\!\lambda\epsilon\omega(a+d)\ \ \mbox{\textrm{\textrm{if }$ a\!+\!d> D_1$}}\\
&\!-\!\epsilon\theta(D_1\!+\!1)\!+\!\psi(a+d)\;, \mbox{ \textrm{and } $d\leq D_a$},\\
&f(a,D_a)\!+\!\epsilon\left(\theta(a\!+\!d)\!-\!\theta(a\!+\!D_a)\right),\  \mbox{\textrm{otherwise}}.\\
\end{aligned}
\right.
\end{align}

We then prove that \eqref{eq:fad} is a valid solution to \eqref{eq:Bellmaneq}.

Since the threshold $D_a$ can be obtained by computing $f(a,d)$, we derive $f(a,d)$ for $d>D_a$ and $d\leq D_a$, separately. 

We first derive $f(a,d)$ for $d>D_a$. For the threshold-based policy $c_D$ with the threshold $D_a$, the optimal action of state $(a,d)$ is to schedule, when $d>D_a$. Then we define $\Delta f(\xi;\rho,\sigma)\triangleq f(\xi,\rho)-f(\xi,\sigma)$. Given $a$, we obtain $\Delta f(1;d,D_{a})$ in \eqref{eq:faddgeqDa1}.
\begin{figure*}[t]
\normalsize
\begin{align}\label{eq:faddgeqDa1}
\Delta f(1;d,D_{a})
=&\e(\theta(d+1)-\theta(D_a+1))+\sum_{k\!=\!1}^{a\!-\!1}
\Big(\lambda(1\!-\!\lambda)^{k\!-\!1}
\big(\Delta f(k;d+a-k,D_{a}+a-k)-\e(\theta(d+a)-\theta(D_a+a))\big)\!\notag\\
&\hspace{0mm}+\!(1\!-\!\lambda)^{a\!-\!1}\big(\Delta f(a;d,D_{a})
-\e\left(\theta(d+a)-\theta(D_a+a)\right)\big)\Big).
\end{align}  
\hrulefill\vspace{-2em}
\end{figure*}
Since \eqref{eq:faddgeqDa1} holds for any $a$, we obtain
\begin{align}\label{eq:faddgeqDa11}
g(1,d)\!=\!\sum_{k\!=\!1}^{a\!-\!1}\lambda(1\!-\!\lambda)^{k\!-\!1}
g(k,a\!+\!d\!-\!k)\!+\!(1\!-\!\lambda)^{a\!-\!1}g(a,d),
\end{align}
where $g(a,d)\!=\!\Delta f(a;d,D_{a})-\e(\theta(a\!+\!d)\!-\!\theta(a\!+\!D_a))$. We find that $g(a,d) = 0$ is a valid solution to \eqref{eq:faddgeqDa11}. 
Thus, we obtain $f(a,d)$ for $d>D_a$ as \eqref{eq:fad}.




We then derive $f(a,d)$ for $d\leq D_a$. For the threshold-based policy $c_D$ with the threshold $D_a$, the optimal action of state $(a,d)$ is to idle, when $d<D_a$. Then, we obtain $f(a,d)$ as
\begin{align}\label{eq:recsolve}
f(a,d) =&-J^{\ast}\!+\!v(d\!+\!a)\!+\!\lbd f(1,d\!+\!a)+\!(1-\lbd)f(a+1,d) \!\notag\\
=&-J^{\ast}\!+\!v(d\!+\!a)\!+\!\lbd f(1,D_1)\notag\\
&+\!\epsilon\lbd(\theta(a\!+\!d\!+\!1)\!-\!\theta(D_1\!+\!1))\!+\!(1-\lbd)f(a+1,d),
\end{align}
for any $a$, $d$, and $k$ satisfying $d\leq D_a$ and $a+d> D_1$. Expanding the last term recursively, we obtain
\begin{align}\label{eq:recsolve1}
  f(a,d)\!= &\sum\limits_{s=0}^{k-1}\!(1-\lbd)^s\left(-J^{\ast}\!+\!v(d\!+\!a\!+\!s)\!\!+\! \lbd f(1,D_1)\right)\notag\\
  &+\!\epsilon\lbd\sum\limits_{s=0}^{k-1}(1-\lbd)^s(\theta(a+d+1+s)-\theta(D_1+1))\notag\\
  &+\!(1\!-\!\lbd)^{k}f(a\!+\!k,d)\notag\\
  =\!   &f(1,D_1)\! -\!\frac{J^{\ast}}{\lbd}\!+\!\lbd\e\omega(a\!+\!d)\!-\!\e\theta(D_1\!+\!1)\!+\!\psi(a\!+\!d)\notag\\
&+\!(1\!-\!\lbd)^{k}\big(f(a\!+\!k,d)\!-\!f(1,D_1)\! +\!\frac{J^{\ast}}{\lbd}\!\notag\\
&-\!\lbd\e\omega(a\!+\!d\!+\!k)\!+\!\e\theta(D_1\!+\!1)\!-\!\psi(a\!+\!d\!+\!k)\big),
\end{align}

From \eqref{eq:recsolve1}, 
we obtain $f(a,d)$ as
\begin{align}\label{eq:fadg}
f(a,d)=&f(1,D_1)-\frac{J^{\ast}}{\lbd}+\lbd\e\omega(a+d)\notag\\
&-\e\theta(D_1+1)+\psi(a+d),
\end{align}
for $a+ d\geq D_1$. In addition, we observe from \eqref{eq:fadg} that
$f(a_1,d_1) = f(a_2,d_2)$ for any $a_1$, $a_2$, $d_1\!<\! D_{a_1}$, and $d_2\!<\! D_{a_2}$. 
By combining \eqref{eq:Bellmaneq} with $f(a_1,d_1) = f(a_2,d_2)$, we obtain
\begin{align}\label{eq:fa0Ja}
f(a,0) 
    &= J^{\ast}-v(a-1)+f(a-1,0),\notag\\
    &=(a-1)J^{\ast}-\sum_{h=1}^{a-1}v(h),
\end{align}
for $a\leq D_1$. We clarify that both actions for the state $(a,D_a)$ are optimal and formulate it as
\begin{align}\label{eq:u0u1}
\mu_0(a,D_a)=\mu_1(a,D_a).   
\end{align}
Based on \eqref{eq:u0u1}, we obtain
\begin{align}
    f(a,D_a)  &= \mu_0(a,D_a)-J^{\ast} \notag\\
    &= m - J^{\ast} + \epsilon\mu_0(a,D_a)+(1-\epsilon)\mu_0(a,0)\notag\\
    & = m + \epsilon f(a,D_a) + (1-\epsilon)f(a,0). 
\end{align}
Hence, we obtain $f(a,D_a)$ as
\begin{align}\label{eq:f1}
f(a,D_a)=\frac{m}{1-\e}+f(a,0).
\end{align}
Moreover, based on $f(1,0)=0$, we obtain
\begin{align}\label{eq:f1D1x}
    f(1,D_1) = \frac{m}{1-\e}+f(1,0)=\frac{m}{1-\e}.
\end{align}
By substituting \eqref{eq:f1D1x} into \eqref{eq:fadg} and combining \eqref{eq:fadg} with \eqref{eq:fa0Ja}, we obtain the solution of $f(a,d)$ for $d<D_a$ as the first and the second cases in \eqref{eq:fad}. 
Furthermore, by combining \eqref{eq:fad} with $\mu_0(\!1,\!D_1\!)\! =\! \mu_1(\!1,\!D_1\!)$, we\! express the optimal AoI, $J^{\ast}$, as a function of the threshold $D_1$ and the service charge $m$, given by
\begin{align}\label{eq:Jmsolution}
J^{\ast}=\frac{1}{D_1}\left(\frac{m}{1-\e}+\sum_{h=1}^{D_1}v(h)\right).
\end{align}

We then show that $f(a,d)$ is non-decreasing with $d$. Since $f(a,d)$ are given for three cases in \eqref{eq:fad}, we first prove that $f(a,d)$ is non-decreasing with $d$ for these three cases separately. We note that $\psi(h)$, $\theta(h)$, and $\omega(h)$ are non-decreasing functions with $h$, since $v(h)$ is a non-decreasing function with $h$. Based on the expression for $f(a,d)$ in \eqref{eq:fad}, we obtain
Based on the expression for $f(a,d)$ in \eqref{eq:fad}, we obtain
\begin{align}\label{eq:23}
    f(a,d_2) - f(a,d_1) = \e(\theta(a+d_2)-\theta(a+d_1))\geq 0,
\end{align}
if $D_a\leq d_1\leq d_2$. Differently, if $d_1\leq d_2 \leq D_a$ and $D_1< a+d_1\leq a+d_2$, we obtain
\begin{align}\label{eq:fadincreases2}
f(a,d_2) \!-\! f(a,d_1)\!= &\lbd\e(\omega(a\!+\!d_2)\!-\!\omega(a+d_1))\notag\\
&+\!(\psi(a\!+\!d_2)\!-\!\psi(a\!+\!d_1))\geq0.
\end{align}
Based on \eqref{eq:23} and \eqref{eq:fadincreases2}, we have proved that $f(a,d)$ is non-decreasing with $d$ for the second case, i.e., $a+d>D_1$ and $d\leq D_a$, and the third case, i.e., $d>D_a$, in \eqref{eq:fad}, respectively. We then prove that $f(a,d)$ is non-decreasing with $d$ for the first case, i.e., $1\leq a+d\leq D_1$, in \eqref{eq:fad}. Since $\mu_0(a,D_a)=\mu_1(a,D_a)$,
by substituting $(a,d) = (1,D_1)$ into \eqref{eq:fadg}, we obtain
\begin{align}\label{eq:vjd1}
    J^{\ast} &= \lbd(\!\lambda\epsilon\omega(D_1+1)\!-\!\epsilon\theta(D_1\!+\!1)\!+\!\psi(D_1+1))\notag\\
    &=\lambda(1-\e)\omega(D_1).
\end{align}
We note that
\begin{align}
    \omega(D_1) &= \frac{\psi(D_1)-\theta(D_1)}{1-\lbd-\e}\notag\\
    &=\frac{\sum\limits_{k=0}^{\infty}((1-\lbd)^k-\e^k)v(D_1+k)}{1-\lbd-\e}\notag\\
    &\geq \frac{\sum\limits_{k=1}^{\infty}((1-\lbd)^k-\e^k)v(D_1+1)}{1-\lbd-\e}\notag\\
    & = \frac{v(D_1+1)}{\lbd(1-\e)}.
\end{align}
Hence, we obtain
\begin{align}
    J^{\ast}=\lambda(1-\e)\omega(D_1)\geq v(D_1+1).
\end{align}
Based on the expression of $f(a,d)$ in \eqref{eq:fad}, we obtain
\begin{align}\label{eq:fadincreases1}
    f(a,d_2)-f(a,d_1) = (d_2-d_1)J^{\ast}-\sum\limits_{a+d_1}^{a+d_2-1}v(h)\geq 0,
\end{align}
if $a+d_1\leq a+d_2\leq D_1$. Hence, we have proved that $f(a,d)$ is non-decreasing function with $d$ for three cases in \eqref{eq:fad}. 

We note that $f(a,d)$ for the first case and the second case in \eqref{eq:fad} are equal to each other when $a+d = D_1+1$, i.e.,
\begin{align}\label{eq:fadincrease3}
&D_1J^{\ast}\!-\!\sum_{h=1}^{D_1}v(h)\notag\\
&=\frac{m}{1\!-\!\epsilon}\!-\!\frac{J^{\ast}}{\lambda}\!+\!\lambda\epsilon\omega(D_1\!+\!1)\!-\!\epsilon\theta(D_1\!+\!1)\!+\!\psi(D_1\!+\!1).
\end{align}
Combining \eqref{eq:fadincreases2}, \eqref{eq:fadincreases1}, with \eqref{eq:fadincrease3}, we obtain that $f(a,d)$ is non-decreasing for $d\leq D_a$.

Moreover, we note that 
\begin{align}\label{eq:fadnondecreaseP1}
    f(a,d_1)\leq f(a,D_a)
\end{align}
and
\begin{align}\label{eq:fadnondecreaseP2}
    f(a,d_2)\! =\! f(a,D_a)\! +\! \epsilon(\theta(a\!+\!d_2)\!-\!\theta(a\!+\!D_a))\geq f(a,D_a),
\end{align}
when $d_1\leq D_a\leq d_2$. Combining \eqref{eq:fadnondecreaseP1}, \eqref{eq:fadnondecreaseP2}, \eqref{eq:23}, with the fact that $f(a,d)$ is non-decreasing for $d\leq D_a$, we obtain that $f(a,d)$ is a non-decreasing function with $d$. 

Based on the cost-to-go function $f(a,d)$ obtained in \eqref{eq:fad}, 
we next derive the threshold $D_{a}$ of the policy $c_{D}$. 
Then, based on \eqref{eq:u0u1}, we express \eqref{eq:u0u1} as
\begin{align}\label{eq:aleqD11}
&\mu_0(a,D_a)-\mu_1(a,D_a)\notag\\
&=-\!m \!+\!(1\!-\!\e)\Big(v(a\!+\!D_a)\!-\!v(a)\!+\!\lambda \Delta f(1;a+D_{a},a)\notag\\
&~~~+\!(1\!-\!\lbd)\Delta f(a+1;D_{a},0)\Big)\notag\\
&=-\!\left(a\!+\!\frac{1}{\lbd}\!-\!1\right)J^{\ast}\!
+\!\sum_{h=1}^{a-1}v(h)\!+\!\lbd\e\omega(a\!+\!D_a)\notag\\
&~~~~-\!\e\theta(D_1\!+\!1)\!+\!\psi(a\!+\!D_a)=0,
\end{align}
when $1\!\leq\! a\!<\! D_1$. By substituting \eqref{eq:Jmsolution} into \eqref{eq:aleqD11}, we obtain \eqref{eq:thresholdD1}. It is noted that since both $\omega(h)$ and $\psi(h)$ monotonically increase with $h$, there is at most one possible number $D_a$ satisfying \eqref{eq:thresholdD1}.

When $a\geq D_{1}$, we express \eqref{eq:u0u1} as
\begin{align}\label{eq:ageqD11}
&\mu_0(a,D_a)-\mu_1(a,D_a)\notag\\
&=-\!m \!+\!(1-\e)\big(v(a\!+\!D_a)\!-\!v(a)\!+\!\lambda\Delta f(1;a+D_{a},a)\notag\\
&~~~+\!(1\!-\!\lbd)\Delta f(a+1;D_{a},0)\big)\notag\\
&=(1\!-\!\e)\left(\psi(a\!+\!D_a)\!-\!\psi(a)
\!+\!\lbd\e\left(\omega(a\!+\!D_a)\!-\!\omega(a)\right)\right)\!-\!m\notag\\
&=0,
\end{align}
which leads to \eqref{eq:thresholdlD2}.

\section{Proof for Theorem \ref{Theorem:2}}\label{Appendix:B}
When $a\leq D_1$, we combine \eqref{eq:Jmsolution} and \eqref{eq:vjd1} as
\begin{align}\label{eq:whittlelD1}
I_v(1,d,\lbd,\e)= (1\!-\!\e)\left(\!\lbd(1\!-\!\e)D_1\omega(D_1)\!-\!\sum_{h=1}^{D_1}v(h) \right).
\end{align}
We note that $D_1=d$ holds for $a=1$. Otherwise, by substituting $D_a=d$ into \eqref{eq:thresholdD1}, the threshold $D_1$ is the minimum positive number satisfying \eqref{eq:WD1calcu}.

When $a>D_1$, we obtain the Whittle index from \eqref{eq:thresholdlD2} as
\begin{align}\label{eq:whittlegD2}
I_v(a,d,\lbd,\e)=(1-\e)\big(&\psi(a+d)\!-\!\psi(a)
\!\notag\\
&+\!\lbd\e\left(\omega(a+d)\!-\!\omega(a)\right)\big).
\end{align}
With \eqref{eq:whittlelD1} and \eqref{eq:whittlegD2}, the Whittle index is obtained as \eqref{eq:Whittle}.
\end{appendices}

\bibliographystyle{IEEEtran} 
\bibliography{bibli}

\end{document}